\documentclass[epsfig]{article}
\usepackage{graphics}

\begin{document}

\title{Single Production of Fourth Family $b^{'}$ Quarks at the Large Hadron electron Collider}

\author{O. \c{C}ak{\i}r\footnote{ocakir@science.ankara.edu.tr} and 
V. \c{C}etinkaya\footnote{cetinkaya@science.ankara.edu.tr}\\
Physics Department, Faculty of Sciences, Ankara University,\\
06100, Tando\u{g}an, Ankara, Turkey}

\maketitle

\begin{abstract}
We examined the single production of
fourth family $b^{\prime}$ quarks at the Large Hadron electron
Collider (LHeC). We have analyzed the background and the signal processes for the mass range
300-700 GeV. We find the discovery region for the optimal 
bounds of $V_{qb^{\prime}}$ matrix elements.
\end{abstract}

\section* {1. Introduction}
The Large Hadron Collider (LHC) with the high centre of
mass energy and high luminosity is expected to give answers to some
questions such as Higgs phenomena and the flavor problem in the
Standard Model (SM) of elementary particles.
In addition to the $pp$ collider, $ep$ collider namely, the Large Hadron electron Collider (LHeC) \cite{Dainton06} could give
 complementary information for the new physics. The LHeC facility will use the proton beam with energy of 7 TeV
from the LHC and the electron beam with energy of 70/140 GeV from a
linear accelerator or storage ring.
 The LHeC would also be expected to have sensitivity to the new quarks and
 leptons.

One of the questions left open by the SM is the replication of
fermion generations. The SM does not provide a mathematical tool
to predict the number of fermion families.
 One of the main aims of the future high energy collider experiments is
 to determine the number of fermion families.
 The discovery of fourth family fermions that have sequential couplings
 could play an important role to understand the flavor structure of the
 SM \cite{Fritzsch87}.
The production of fourth family fermions have been studied at hadron colliders \cite{Kribs07}, 
lepton colliders \cite{Ciftci05} and ep collider \cite{Cakir09}.

In this work, we investigate the discovery potential of the LHeC for
the single production of fourth family $b^{\prime}$ ($\bar{b^{\prime}}$)
quarks via the process 
$e^{-}p\rightarrow b^{\prime}\nu_{e}$
($e^{+}p\rightarrow\bar{b^{\prime}}\bar{\nu_{e}}$). We calculate the
cross sections of signal and corresponding backgrounds.
 The decay widths and branching ratios of the fourth family $b^{\prime}$ quarks are calculated in the mass range 300-800 GeV.
 For the numerical calculations, we have implemented the new interaction vertices into the CompHEP \cite{Pukhov99}
package and used the parton distribution function (PDF) CTEQ6M \cite{Pumplin07}.

\section*{2. Single Production and Decay of $b^{\prime}$ Quark}

The simplest extension of the SM is to add a sequential fourth
family fermions. Here, the left-handed components transform as a doublet of $SU(2)_{L}$ and right-handed components as singlets.
 The fourth family $b^{\prime}$ quark interacts with the quarks $q_{i}$ via the exchange of SM gauge bosons $(\gamma,g,Z^{0},W^{\pm})$.
 The interaction lagrangian is given by

\begin{eqnarray}
L=-g_{e}Q_{b^{\prime}}\bar{b^{\prime}}\gamma^{\mu}b^{\prime}A_{\mu}-g_{s}\bar{b^{\prime}}T^{a}\gamma^{\mu}b^{\prime}G_{\mu}^{a}-\frac{g}{2cos\theta_{W}}\bar{b^{\prime}}\gamma^{\mu}\left(g_{V}-g_{A}\gamma^{5}\right)b^{\prime}Z_{\mu}^{0}
\nonumber \\
-\frac{g}{2\sqrt{2}}V_{q_i b^{\prime}}\bar{b^{\prime}}\gamma^{\mu}\left(1-\gamma^{5}\right)q_{i}W_{\mu}^{-}+h.c.
\end{eqnarray}

\begin{figure}
\includegraphics{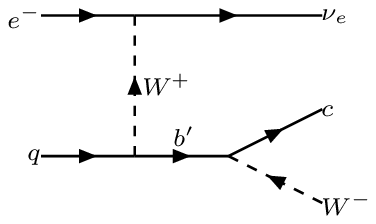}\includegraphics{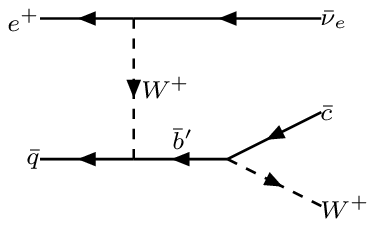}
\caption{The diagrams for the single production of $b^{\prime}(\bar{b^{\prime}})$ quark at $ep$ collision.\label{fig1}}
\end{figure}

\begin{table}
\caption{The total decay widths and the branching ratios of
$b^{\prime}$ quark depending on its mass. \label{tab1}}
\begin{tabular}{ccccc}
\hline
Mass (GeV) & $\Gamma$ (GeV) & $W^{-}c$($\%$) & $W^{-}t$($\%$) & $W^{-}u$($\%$) \\
\hline
300 & 0.18 & 66 & 30 & 3.9 \\
400 & 0.55 & 52 & 45 & 3.0 \\
500 & 1.22 & 46 & 52 & 2.7 \\
600 & 2.25 & 43 & 55 & 2.5 \\
700 & 3.72 & 41 & 57 & 2.4 \\
800 & 5.70 & 40 & 58 & 2.3 \\
\hline
\end{tabular}
\end{table}

\begin{figure}[tbp]
\includegraphics{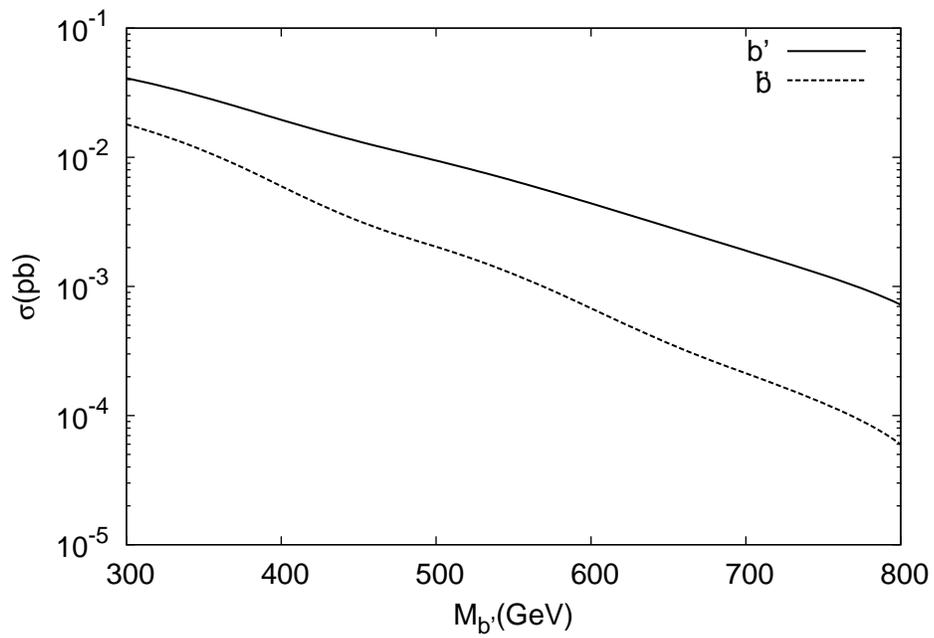}
\caption{The total cross section for the process
$e^{-}p\rightarrow b^{\prime}\nu_{e}$ (solid line) and
$e^{+}p\rightarrow\bar{b^{\prime}}\bar{\nu_{e}}$ (dashed line)
with $\sqrt{s}=1.4$ TeV. \label{fig2}}
\end{figure}

where $g_{e}$, $g$ are electro-weak coupling constants, and
$g_{s}$ is the strong coupling constant. The vector fields $A_{\mu}$, $G_{\mu}$,
$Z_{\mu}^{0}$ and $W_{\mu}^{\pm}$ denote photon,
gluon, $Z^{0}$-boson and $W^{\pm}$-boson, respectively.
$Q_{b^{\prime}}$ is the electric charge of fourth family 
$b^{\prime}$ quark; $T^{a}$ are the Gell-Mann matrices. 
The vector and axial-vector type couplings $g_{V}$ and $g_{A}$ of neutral
weak current are defined as in the SM. Finally, the
$V_{qb^{\prime}}$ denotes the elements of extended 4 $\times$ 4
CKM mixing matrix which are constrained by flavor physics.
We use the parametrization, which is well motivated in the recent studies \cite{Hou07}:
$V_{ub^{\prime}}=0.028$, $V_{cb^{\prime}}=0.116$,
 $V_{tb^{\prime}}=0.15$, $V_{t^{\prime}b^{\prime}}=0.99$, and 
we assume $m_{b^{\prime}}<m_{t^{\prime}}$ with a mass splitting of $m_{t^{'}}-m_{b^{'}}\approx 50$ GeV. 

The collider detector at Fermilab (CDF) has already constrained the masses of fourth-family quarks: 
$m_{t^{\prime}}>311$ GeV at $95\%$ CL. \cite{CDF08}, and $m_{b^{\prime}}>326$ GeV at $95\%$ CL. \cite{CDF09} 
However, there seems to be some parameter space (mass, mixing angles) left for the fourth-family 
quarks which could be investigated in the future experiments.

\begin{figure}
\includegraphics{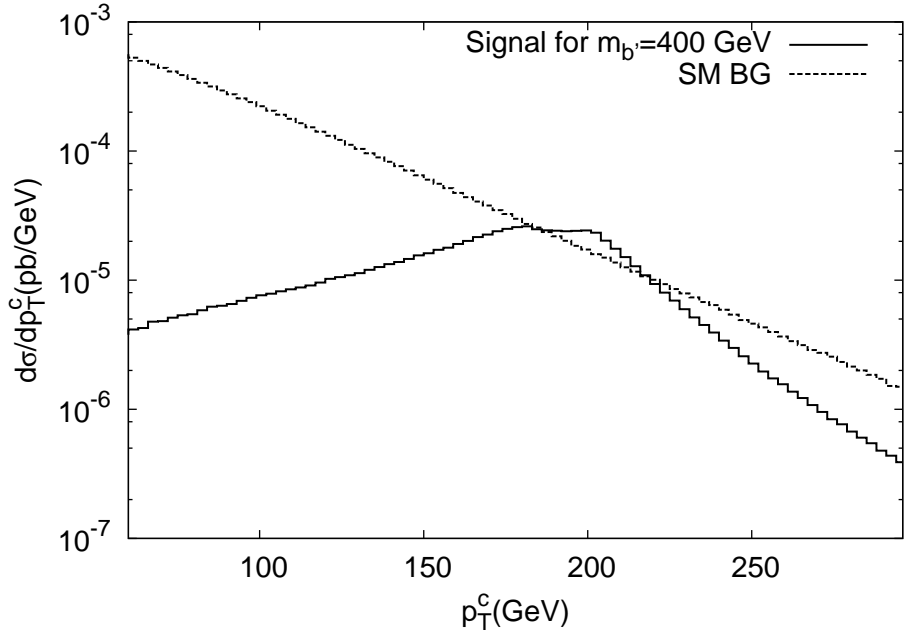}
\caption{Transverse momentum distribution of the final state charm quark for the subprocess
$e^{+}p\rightarrow W^{+}\bar{c}\bar{\nu_{e}}$. The solid line is for the signal with $m_{b^{\prime}}=400$ GeV and
dashed-line corresponds to the SM background. \label{fig3}}
\end{figure}

The relevant diagrams for the single production of $b^{\prime}(\bar{b^{\prime}})$ quark and their subsequent decays
are shown in Fig. \ref{fig1}. The total decay widths and the
branching ratios of $b^{\prime}$ quark within the SM framework are
presented in Table \ref{tab1}. The branchings to $Wc$ remains dominant for the $b'$ mass range of 300-450 GeV, 
while the $Wt$ channel becomes more pronounced for the high mass region (450-800 GeV). In Fig. \ref{fig2}, 
we show the cross sections for the single production of the fourth generation $b^{\prime}$ and $\bar{b^{\prime}}$ quarks
depending on their masses at the LHeC with $\sqrt{s}=1.4$ TeV.
The cross-sections for $b^{\prime}$ and $\bar{b^{\prime}}$ single production 
can be added to enrich the statistics for the analysis.

\begin{figure}
\includegraphics{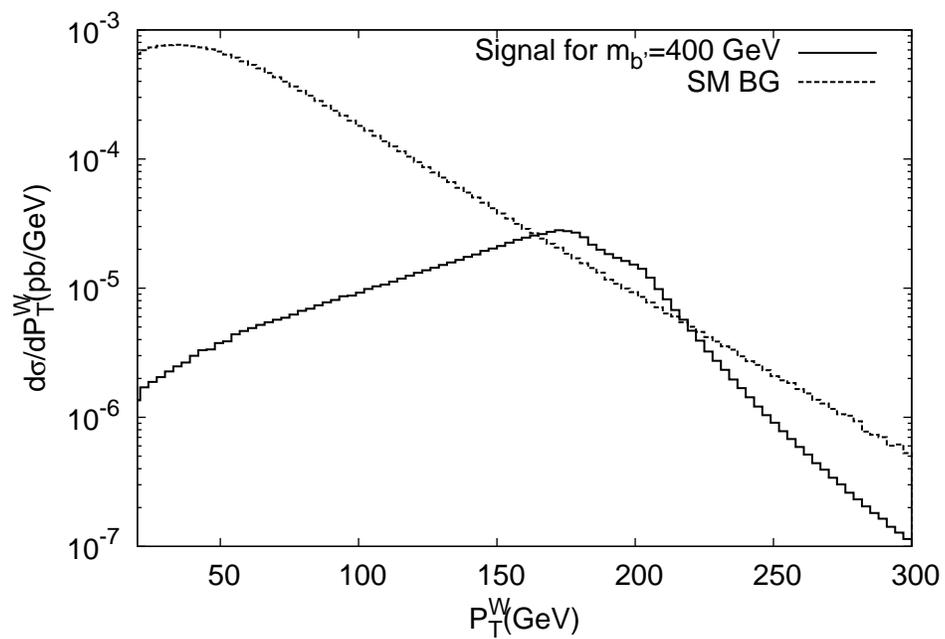}
\caption{The transverse momentum distribution of the final state $W^{+}$-boson for the
subprocess $e^{+}p\rightarrow W^{+}\bar{c}\bar{\nu_{e}}$. The
solid line and dashed line correspond to the SM background and
the signal for $m_{b^{\prime}}=400$ GeV, respectively. \label{fig4}}
\end{figure}

\begin{figure}
\includegraphics{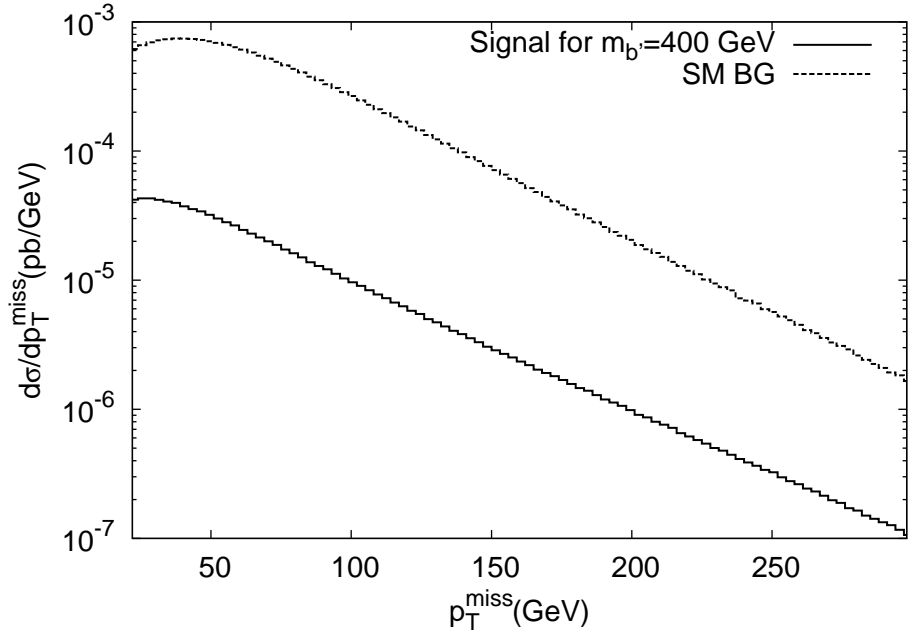}
\caption{The missing transverse momentum distribution of the final state neutrino for the
subprocess $e^{+}p\rightarrow W^{+}\bar{c}\bar{\nu_{e}}$. The
solid line and dashed line show the signal and background, respectively. \label{fig5}}
\end{figure}

In Fig. \ref{fig3}, we display the transverse momentum distributions of the
final state c-quark for the signal and background. 
The $p_{T}$ distribution shows a peak around 200 GeV for the 
$b^{\prime}$ signal with mass $m_{b^{\prime}}=400$ GeV. We can compare this
distribution with that of the corresponding background and 
we apply a $p_{T}$ cut to reduce the background.
We also plot the $p_{T}$ distributions of $W^{+}$-boson and missing
$P_{T}$ as shown in Figs. \ref{fig4} and \ref{fig5}. 
We see that the following transverse momentum cuts are 
required for the analysis: $p_{T}^c>20$ GeV, $p_{T}^{\ell}>20$ GeV and $p_{T}^{miss}>20$ GeV. 
The signal will have a charm quark in the final state, 
and the charm quark hadronizes immediately after it is 
produced. A charmed jet has a secondary vertex mass around 1 GeV. 
Here, we assume one can have much success for tagging the charm hadrons in the future 
experimental environment.

\begin{figure}
\includegraphics{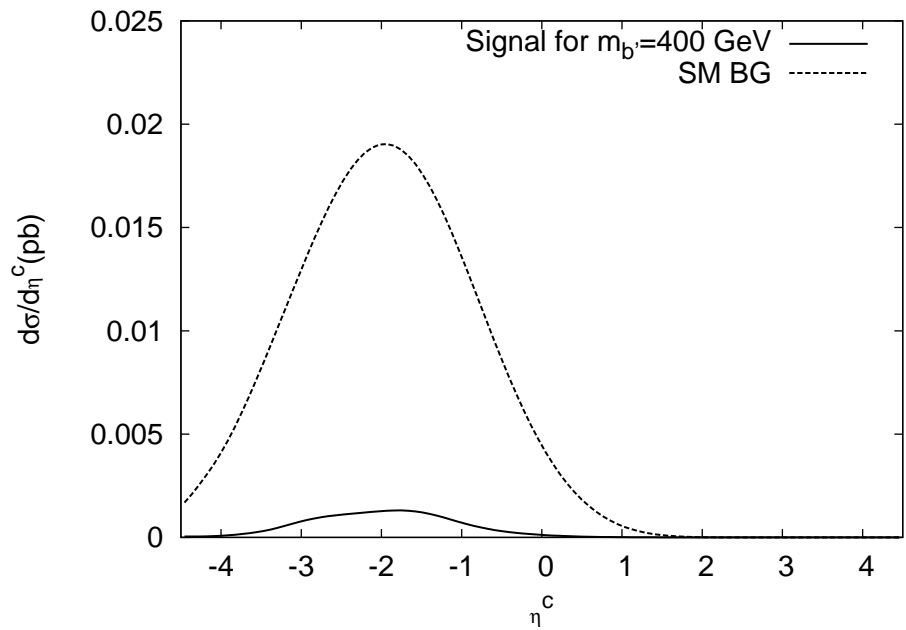}
\caption{The rapidity distribution of c quark from the
subprocess $e^{+}p\rightarrow W^{+}\bar{c}\bar{\nu_{e}}$. 
The solid line shows the signal with $m_{b^{\prime}}=400$ GeV and dashed line shows the SM background. \label{fig6}}
\end{figure}

\begin{figure}
\includegraphics{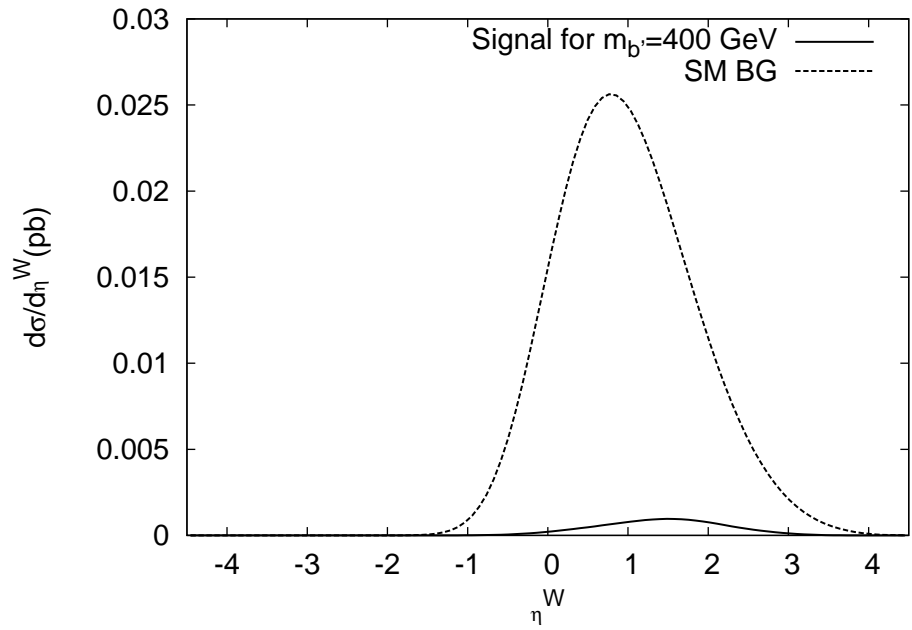}
\caption{The rapidity distribution of final state $W^{+}$-boson from
the subprocess $e^{+}p\rightarrow W^{+}\bar{c}\bar{\nu_{e}}$. 
Solid line is for the signal with $m_{b^{\prime}}=400$ GeV and 
dashed line is for the SM background. \label{fig7}}
\end{figure}

\begin{figure}
\includegraphics{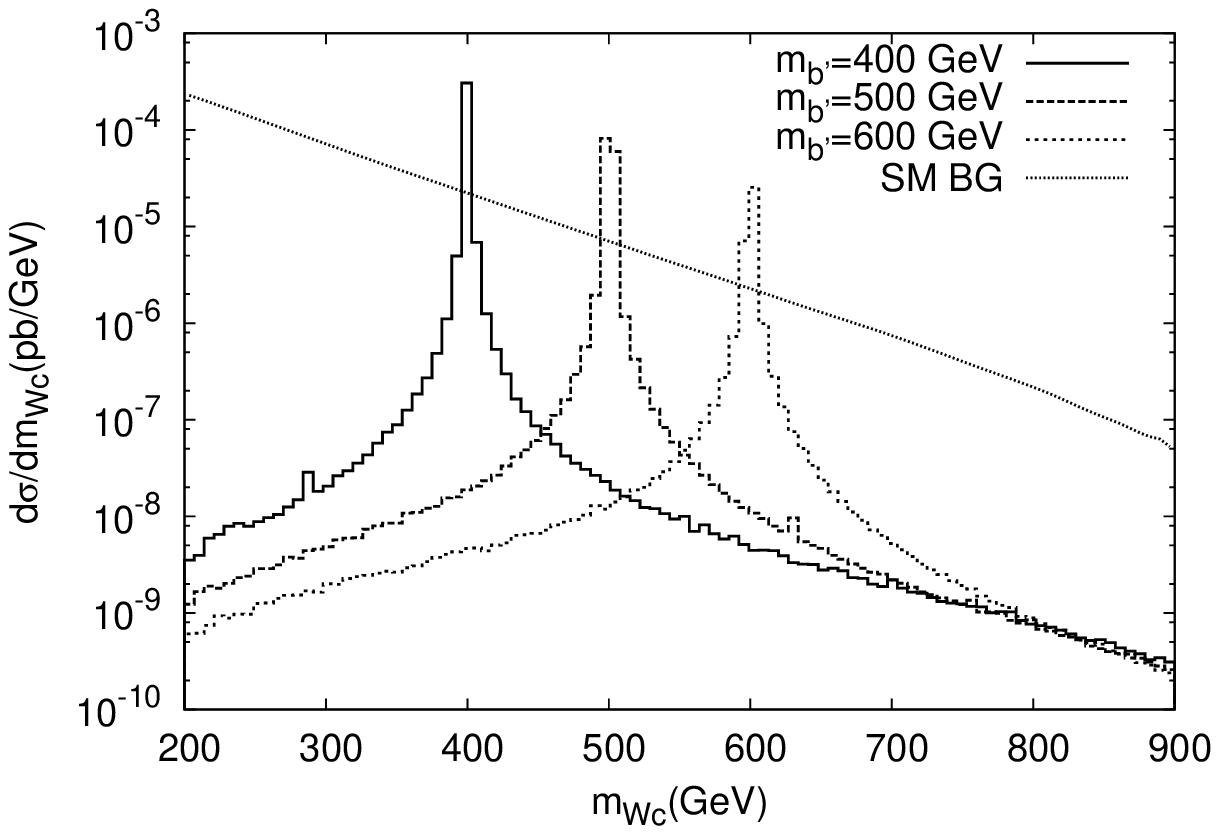}
\caption{The invariant mass distribution of the $Wc$ system for the signal and background. \label{fig8}}
\end{figure}

The rapidity for the signal and background show different distributions as seen in Figs. \ref{fig6} and \ref{fig7}. 
Fig. \ref{fig8} shows the invariant mass of $Wc$
system. The peaks in the invariant mass distribution show the $b^{\prime}$ signal with the masses of
400, 500 and 600 GeV.

\section* {3. Analysis}
We calculate the number of events in the invariant mass interval around 
the $m_{b^{\prime}}$, namely $|m_{b^{'}}-m_{Wc}|<10-20$ GeV according to 
the mass and the decay width of
$b^{\prime}$ to obtain the visible signal over the background. 
In this case, a significant reduction on the background can be obtained.
In the following equation, $\sigma_{S}$ and $\sigma_{B}$ denote signal and background 
cross sections in the selected mass bins. Assuming the Poisson statistics, the estimations for the statistical 
significance ($SS$) of signal is obtained by assuming an integrated luminosity of $L_{int}=10^{4}$pb$^{-1}/$year

\begin{equation}
SS=\sqrt{2L_{int}\epsilon\left[\left(\sigma_{S}+\sigma_{B}\right)ln\left(1+\sigma_{S}/\sigma_{B}\right)-\sigma_{S}\right]}.
\end{equation}

The number of events for the signal and background processes $e^{+}p\rightarrow W^{+}\bar{c}\bar{\nu_{e}}$ (
$e^{-}p\rightarrow W^{-}c\nu_{e}$ ) are calculated as $N_{S}=\sigma_{S}\epsilon L_{int}$, 
where we take into account $W^{\pm}$ leptonic decays. 
In Table \ref{tab2}, the results for the signal significance of the $b^{\prime}$ quark are shown for 
 $L_{int}=10^{4}pb^{-1}$. Here, we assume the c-tagging efficiency as $\epsilon=30\%$. 
The single production of $b^{\prime}$ quark can be observed at the LHeC in the mass 
range of 300-700 GeV provided the current bounds on the mixings with the other families are present.

\section*{4. Conclusion}

The LHC can discover the fourth family quarks in pairs and measure their masses with a good accuracy. 
We have explored the potential of the LHeC for searching for the $b^{\prime}$ single production in the allowed parameter space. 
If the $b^{\prime}$ quark has large mixing with the other families of the SM, it
can also be produced singly at the LHeC with large numbers. From the single production, 
a unique measurement can be performed for the family mixings with the four families. 

\begin{table}
\caption{The total cross section for the signal ($\sigma_{S}$) and
background ($\sigma_{B}$). The statistical significance ($SS$) values calculated for the process $e^{+}p\rightarrow W^{+}\bar{c}\bar{\nu_{e}}$ ($e^{-}p\rightarrow W^{-}c\nu_{e}$) at LHeC with $\sqrt{s}=1.4$ TeV and
$L_{int}=10^{4}pb^{-1}$. \label{tab2}}
\begin{tabular}{cccc}
\hline
$m_{b^{\prime}}$ (GeV) & ${\sigma_{S}}$ (pb) & ${\sigma_{B}}$ (pb) & $SS$ \\
\hline
300 & $2.07\times 10^{-2}$($8.38\times 10^{-3}$) & $2.86\times 10^{-3}$($2.87\times 10^{-3}$) & $13.2$($6.48$) \\
400 & $7.85\times 10^{-2}$($2.24\times 10^{-3}$) & $8.87\times 10^{-4}$($8.82\times 10^{-4}$) & $8.53$($3.2$) \\
500 & $3.33\times 10^{-3}$($6.84\times 10^{-4}$) & $2.82\times 10^{-4}$($2.84\times 10^{-4}$) & $5.94$($1.74$) \\
600 & $1.42\times 10^{-3}$($2.09\times 10^{-4}$) & $9.10\times 10^{-5}$($9.09\times 10^{-5}$) & $4.12$($0.95$) \\
700 & $5.65\times 10^{-4}$($6.11\times 10^{-5}$) & $2.87\times 10^{-5}$($2.84\times 10^{-5}$) & $2.72$($0.5$) \\
\hline
\end{tabular}
\end{table}


\begin{thebibliography}{20}

\bibitem{Dainton06} J. B. Dainton, M. Klein, P. Newman, E. Perez and F. Willeke, JINST {\bf 1}, 10001 (2006). 
\bibitem{Fritzsch87} H. Fritzsch, Phys. Lett. B {\bf 184}, 391 (1987); H. Fritzsch, Phys. Lett. B {\bf 289}, 92 (1992); 
A. Datta, Pramana {\bf 40}, L503 (1993); A. Celikel, A. K. Ciftci, and S. Sultansoy, Phys. Lett. B {\bf 342}, 257 (1995); 
B. Holdom, JHEP {\bf 0708}, 069 (2007).
\bibitem{Kribs07} E. Arik et al., Phys. Rev. D {\bf 58}, 117701 (1998); 
I.F. Ginzburg, I.P. Ivanov, A. Schiller, Phys. Rev. D {\bf 60}, 095001 (1999);
B. Holdom, JHEP {\bf 0608}, 076 (2006); 
G. D. Kribs, T. Plehn, M. Spannowsky and T. M. P. Tait, Phys. Rev. D {\bf 76}, 075016 (2007); 
V. E. Ozcan, S. Sultansoy and G. Unel, Eur. Phys. J. C {\bf 57} , 621 (2008); 
O. Cakir, H. Duran Yildiz, R. Mehdiyev and I. Turk Cakir, Eur. Phys. J. C {\bf 56}, 537 (2008); 
R. Ciftci, Phys. Rev. D {\bf 78}, 075018 (2008); 
E.V. Ozcan, S. Sultansoy, G. Unel, J. Phys. G {\bf 36}, 095002 (2009);
I.T. Cakir et al., Phys. Rev. D {\bf 80}, 095009 (2009).
\bibitem{Ciftci05} A.K. Ciftci, R. Ciftci, and S. Sultansoy, Phys. Rev. D {\bf 72} , 053006 (2005).
\bibitem{Cakir09}  O. Cakir, A. Senol and A. T. Tasci, Eur. Phys. Lett. {\bf 88}, 11002 (2009); arXiv:0905.4347v1.
\bibitem{Pukhov99} A.Pukhov \emph{et al}.,A. Pukhov et al., arXiv:hep-ph/9908288.
\bibitem{Pumplin07}  J. Pumplin \emph{et al.}, JHEP {\bf 0207}, 012 (2002), arXiv:hep-ph/0201195.
\bibitem{Hou07} W.S. Hou, M. Nagashima, a. Soddu, Phys. Rev. D \textbf{76}, 016004 (2007); M.S. Chanowitz, Phys. Rev. D \textbf{79}, 113008 (2009).
\bibitem{CDF08} CDF Public Note, http://www-cdf.fnal.gov/physics/new/top/2008/tprop\\
/Tprime2.8/public.html.
\bibitem{CDF09} CDF Collaboration, Report No. CDF/PHYS/EXO/PUBLIC/9759 (2009); CDF Collaboration, CDF Note 8495 (2007).

\end{thebibliography}
\end{document}